\documentclass[prl,reprint,superscriptaddress]{revtex4-1}
\usepackage[T1]{fontenc}
\usepackage{tikz}
\usetikzlibrary{bending}
\let\olddot=\dot \usepackage{amsmath,amsthm,amsfonts} \let\dot=\olddot
\usepackage{bbm}
\usepackage[hidelinks]{hyperref}
\def\indic#1{\mathbbm{1}_{\{#1\}}}
\def\P{\mathbb P}
\def\diffd{\text d}

\newtheorem{df}{Definition}

\tikzset{part/.style={fill,circle,inner sep=.4ex,outer sep=.2ex}}

\def\tikzbranch#1#2{\tikz[baseline=-.6ex]{\path node[part] (a){}
	(.4,.15) node[part](b){}
	(.4,-.15) node[part](c){};
\draw[->] (a) -- (b) node[anchor=base west,yshift=-.6ex]{#1};
\draw[->] (a) -- (c) node[anchor=base west,yshift=-.6ex]{#2};}}

\def\tikzpart{\tikz[baseline=-.6ex]\path node[part]{};}
\def\tikzright{\tikz[baseline=.15ex]{\draw (0,1.5ex)--(0,0)--(1.5ex,0)
			   (1.5ex,1.5ex)-- (1.5ex,0) --(3ex,0)
			   (3ex,1.5ex)--(3ex,0)--(4.5ex,0)--(4.5ex,1.5ex);
		\path (2.25ex,.75ex) node[part](a){};
		\draw[->] (a) to[out=70,in=110,looseness=2] (3.75ex,1.2ex);}}
\def\tikzleft{\tikz[baseline=.15ex]{\draw (0,1.5ex)--(0,0)--(1.5ex,0)
			   (1.5ex,1.5ex)-- (1.5ex,0) --(3ex,0)
			   (3ex,1.5ex)--(3ex,0)--(4.5ex,0)--(4.5ex,1.5ex);
		\path (2.25ex,.75ex) node[part](a){};
		\draw[->] (a) to[out=110,in=70,looseness=2] (.75ex,1.2ex);}}
\def\et{\;;\,}

\begin{document}

\title{How to generate the tip of branching random walks evolved to large
  times}
\date{\today}

\author{\'Eric Brunet}
\email{eric.brunet@ens.fr}
\affiliation{Laboratoire de Physique de l'\'Ecole normale sup\'erieure,
ENS, Universit\'e PSL, CNRS, Sorbonne Universit\'e, Universit\'e de Paris,
F-75005 Paris, France}

\author{Anh Dung Le}
\email{dung.le@polytechnique.edu}
\affiliation{CPHT, CNRS, \'Ecole polytechnique, IP Paris,
    F-91128 Palaiseau, France}

\author{Alfred H. Mueller}
\email{amh@phys.columbia.edu}
\affiliation{Department of Physics, Columbia University,
  New York, NY 10027, USA}

\author{St\'ephane Munier}
\email{stephane.munier@polytechnique.edu}
\affiliation{CPHT, CNRS, \'Ecole polytechnique, IP Paris,
    F-91128 Palaiseau, France}

\begin{abstract}
  In a branching process, the number of particles increases exponentially
  with time, which makes numerical simulations for large times
  difficult. In many applications, however,
  only the region close to the extremal
  particles is relevant (the ``tip''). We present a simple algorithm
  which allows to simulate a branching random walk
  in one dimension, keeping only the particles that arrive
  within some distance 
  of the rightmost particle at a predefined time $T$.
  The complexity of the algorithm grows linearly with $T$.
  We can furthermore choose to require that the
  realizations have their rightmost particle arbitrarily far on the right
  from its typical position.
  We illustrate our algorithm by evaluating
  an observable for which no other practical method is known.
\end{abstract}

\maketitle

\section{Introduction}

The branching Brownian motion (BBM) \cite{IkedaNagasawaWatanabe.1968}
and branching random walks (BRW) are
stochastic processes~\cite{VanKampen.2011}
describing the time evolution of increasingly many particles
characterized by their spatial positions~\cite{Bovier.2017}.
These processes, supplemented or not by some selection
mechanism~\cite{brunet:tel-01417420},
can model a range of phenomena in different fields of science, including
physics~\cite{DerridaSpohn.1988},
biology~\cite{Murray.2002} and
chemistry~\cite{AronsonWeinberger.1975},
computer science~\cite{MajumdarDeanKrapivsky.2005}, and even
economics~\cite{BenhabibBrunetHager.2020}.

In many applications of  these branching processes in one space dimension,
it is important to characterize the ``tip'' of
the process, i.e.\@ 
the distribution of particles close to the rightmost particle, in typical and in rare
events~\cite{BrunetDerrida.2011,AidekonBerestyckiBrunetShi.2012,Hallatschek.2011}.
Many properties of the tip can be deduced from solutions to
nonlinear evolution equations;
in the case of the BBM,
the relevant equation \cite{McKean.1975} is a partial
differential equation named after Fisher, Kolmogorov, Petrovsky, Piscounov
\cite{Fisher.1937,KPP.1937}
(FKPP):
\begin{equation}
\partial_t u = \tfrac12\partial_x^2 u + u-u^2.
\label{FKPP}
\end{equation}
For instance, the quantity $\P(R_t\ge x)$, with $R_t$ the position of the
rightmost particle at time $t$, is given by the solution
to \eqref{FKPP} for a step initial condition $u(0,x)=\indic{x\le0}$.
More sophisticated observables can also be expressed;
for example, if $n$ is the number of particles
at time $T$ on the right of $R_T-a$,
then one can show \cite{BrunetDerrida.2011} that $\langle e^{-\lambda
n}\rangle= 1-\int\diffd x\,\partial_a u(T,x)$ where $u(t,x)$ is the
solution to \eqref{FKPP} with initial condition $u(0,x)=
\indic{x\le0}+(1-e^{-\lambda})\indic {x\in(0,a]}$.

This method has however some limitations: obtaining the tail
of the distribution of $n$ is impractical; other
observables, such as the genealogical tree of the rightmost particles
\cite{DerridaMottishaw.2016} cannot be obtained in this way.

Because the number of particles in a branching process increases
exponentially fast with time, direct
Monte Carlo simulations are ill-suited except for small times.
Furthermore, they would not allow
 to study
rare events in which the
rightmost particle sits at a position very different from its expected
position.

In this Letter, we present
an algorithm designed to only generate the particles ending
in an interval $[X-\Delta,+\infty)$ at time $T$, in unbiased
realizations conditioned to either possess at
least one particle to the right of $X$, or to have their rightmost
particle at position $X$ exactly.
This algorithm allows to study the properties of the tip
in rare or typical realizations evolved to very large times.

While reminiscent of the classical spinal decomposition of the
BBM \cite{HardyHarris.2009}, our method is actually quite different as 
we follow a tree of ``special'' particles.

We shall start by exposing the algorithm in the case of a particular BRW
which is easy to implement numerically.
In a second section, we provide a formulation for the BBM.
A third section exposes briefly a variant of our algorithm, and
we conclude by showing the numerical calculation of an observable 
for which the algorithm is especially efficient.

\section{Generating realizations of a BRW with a particle beyond a given point}
\label{genBRW}

We consider a branching random walk (BRW) on a spatial lattice with step
$\delta x$, and
in discrete time with step $\delta t$. The system starts at time $t=0$ with
one single particle at the origin. During each time step, a particle in the
system evolves with the following rules: it can
\begin{equation*}\begin{cases}
\text{\tikzright\quad jump from $x$ to $x+\delta x$}&\text{proba }p_r,\\
\text{\tikzleft\quad jump from $x$ to $x-\delta x$} &\text{proba }p_l,\\
\text{\tikzbranch{}{}\quad duplicate without moving}&\text{proba }r,
\end{cases}\end{equation*}
with $p_l+p_r+r=1$. When a particle duplicates (or branches), it is
replaced by two particles at the same position which evolve independently
afterwards.

We let $R_t$ be the position of the rightmost particle at time $t$, and we
introduce $u(t,x) = \P(R_t\ge x)$.
The probability $u$ satisfies the following 
discretization of \eqref{FKPP}:
\begin{multline}
u(t+\delta  t,x) = p_r u(t,x-\delta  x) + p_l u(t,x+\delta x) \\
+r u(t,x)[2-u(t,x)],
\label{FKPP2}
\end{multline}
with initial condition 
$ u(0,x)=\indic{x\le0}$.

From standard results on FKPP \cite{Bramson.1983},
$u$ evolves at large time as a front centered
around position $m_t=\langle R_t\rangle$.
(We use $\langle\,\cdot\,\rangle$ to denote expectations.)
When $t$ is large, $m_t = v_0 t -\frac{3}{2\gamma_0}\log t +\text{const.}+o(1)$ 
where $v_0$ and $\gamma_0$ are given by
$v_0=v(\gamma_0)=\min_{\gamma} v(\gamma)$, with
$v(\gamma)=\frac1{\gamma\,\delta t}\log\big(
p_r e^{\gamma\,\delta x}+p_l e^{-\gamma\,\delta x}+2r\big);
$ for a
review see~\cite{vanSaarloos.2003}.

Pick a time horizon $T$ and a target $X$.
We introduce ``red particles'' in the BRW in the following way:
\begin{df}
A particle is red if its rightmost offspring at time $T$ lies in $[X,\infty).$
\end{df}
A first goal of the algorithm is to follow the trajectories of all the red
particles in the BRW conditioned on the event that the initial particle is
red. The algorithm works for typical realizations if $X-m_T=\mathcal
O(1)$ or for rare events if $X-m_T\gg 1$.

Introduce $U(t,x)$ as the probability that a given particle at $(t,x)$ is
red. By definition of $u$:
\begin{equation}\label{defU}
U(t,x):=\P\big(\tikzpart\text{ is red}\big)=u(T-t,X-x).
\end{equation}

The probability that a particle at $(t,x)$ is
red and jumps to the lattice site on its right is
\begin{equation*}
\P\big(\tikzright\et \tikzpart\text{ is red}\big)
=p_r U(t+\delta t,x+\delta x).
\end{equation*}
(We write $\P(A\et B)$ to mean $\P(A\text{ and }B)$.) Then,
the probability that the particle at $(t,x)$ jumps right given that it is red
can be written as
\begin{equation}\label{redjr}
\P\big(\tikzright\;\big|\; \tikzpart\text{ is red}\big)
= p_r \frac{U(t+\delta t,x+\delta x)}{U(t,x)}.
\end{equation}
(We write 
$\P(A\;|\;B)=\P(A\et B)/\P(B)$ for the conditional probability of $A$ given
that $B$ is realized.)
Similarly, the conditional probability of jumping left is
\begin{equation}\label{redjl}
\P\big(\tikzleft\;\big|\; \tikzpart\text{ is red}\big)
= p_l \frac{U(t+\delta t,x-\delta x)}{U(t,x)}.
\end{equation}
We now turn to branching. Consider a particle branching at $(t,x)$, being
thus replaced by two children at $(t+\delta t,x)$.
The probability that both these children
are red is $U(t+\delta t,x)^2$ and the probability that exactly one of them is
red is $2U(t+\delta t,x)[1-U(t+\delta t,x)]$. Thus,
the probability to be red and branch into two red is
\begin{equation*}
\P\big(\tikzbranch{red}{red}\et \tikzpart\text{ is red}\big)
= r U(t+\delta t,x)^2,
\end{equation*}
and the conditional probability is
\begin{equation}\label{redbrr}
\P\big(\tikzbranch{red}{red}\;\big|\; \tikzpart\text{ is red}\big)
= r \frac{U(t+\delta t,x)^2}{U(t,x)}.
\end{equation}
Similarly, the conditional probability, given that it is red, that
a particle branches into one red and one non-red is
\begin{multline}\label{redbr}
\P\big(\tikzbranch{red}{non-red}\;\big|\; \tikzpart\text{ is red}\big)
\\
= r \frac{2U(t+\delta t,x)\big[1-U(t+\delta t,x)\big]}{U(t,x)}.
\end{multline}
One checks with \eqref{FKPP2} that the sum of the conditional
probabilities in \eqref{redjr}, \eqref{redjl}, \eqref{redbrr} and
\eqref{redbr} is 1. With these equations, it is possible to generate
realizations of the trajectories of all the red particles
given that the initial particle is red:
 we simply start with a red particle at
$x=0$; then, any red particle can either jump right or left with
probabilities \eqref{redjr} and \eqref{redjl}, branch
into two red with probability \eqref{redbrr} or do nothing with probability
\eqref{redbr}. (In the latter case, the particle is actually branching into
a red particle and a non-red particle that we ignore.) The price to be paid
is that the probabilities of the different events are now time- and space-
dependent.

\medskip

The mechanism can be extended to furthermore follow the trajectories of all the
particles arriving in $[X-\Delta,X)$ for some length $\Delta$. Introduce orange and
blue particles:
\begin{df} A particle is orange if its rightmost offspring at time $T$ lies
in $[X-\Delta,X)$.
\end{df}
\begin{df} A particle is blue if its rightmost offspring at time $T$ lies
in $(-\infty,X-\Delta)$.
\end{df}
\noindent(Then, all non-red particles are either orange or blue.)
Introduce $V_\Delta(t,x)$ as the probability that a particle at $(t,x)$ is
orange. By definition of $u$ and $U$, one has
$$V_\Delta(t,x) 
:=\P\big(\tikzpart\text{ is orange}\big)
= U(t,x+\Delta) -U(t,x).$$
An orange particle can be created by the branching of a red particle: we
replace \eqref{redbr} by the probability for a red to branch into a red and
an orange
\begin{equation}\label{redbro}
\P\big(\tikzbranch{red}{orange}\;\big|\; \tikzpart\text{ is red}\big)
= r \frac{2U(t+\delta t,x)V_\Delta(t+\delta t,x)}{U(t,x)},
\end{equation}
and the probability to branch into a red and a blue
\begin{multline}\label{redbrb}
\P\big(\tikzbranch{red}{blue}\;\big|\; \tikzpart\text{ is red}\big)
\\
= r \frac{2U(t+\delta t,x)\big[1-U(t+\delta t,x+\Delta)\big]}{U(t,x)}.
\end{multline}
Once orange particles are created, we need to follow their trajectories.
Conditioned on the event that a particle is orange, the probabilities that
it
jumps right, jumps left or branches into two orange particles are given
respectively, by \eqref{redjr},
\eqref{redjl} and \eqref{redbrr} with $U$ replaced by $V_\Delta$.
The probability that an orange branches into one orange and one blue
is, similarly to \eqref{redbrb}, $r\times
2 V_{\Delta}(t+\delta t,x)[1-U(t+\delta t,x+\Delta)]/V_\Delta(t,x)$.

To implement our algorithm, we represent
the state of the system at a given time $t$  by two arrays
indexed by $x$
containing the numbers of red and orange particles. To forward
the system to time $t+\delta t$, one
observes that on each site in each set, the numbers of particles
undergoing the different possible events obey multinomial laws with
parameters that we can compute from $u(t,x)$.
This requires to integrate numerically (\ref{FKPP2})
 before the event generation begins.

\begin{figure}[ht]
\includegraphics[width=\columnwidth]{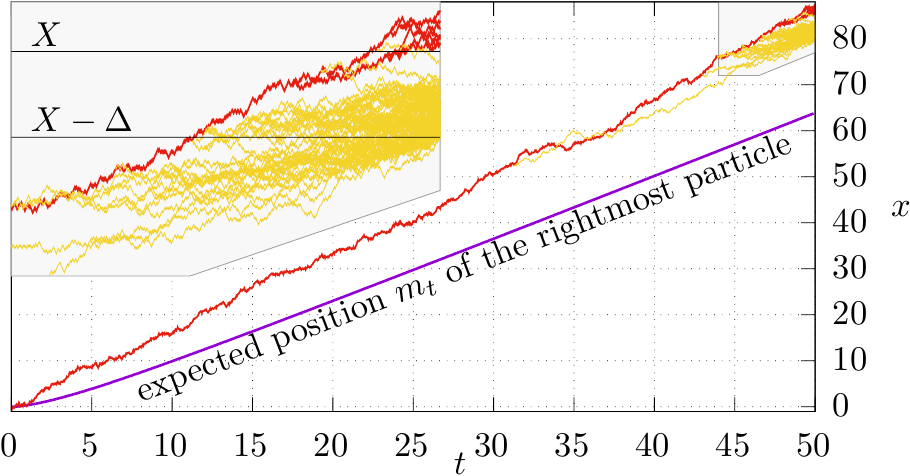}
\caption{\label{fig:realization}
  A realization of the red (dark) and orange (light) particles in
  the BRW up to $T=50$ with $X=85.1\simeq m_T+3\sqrt T$ and $\Delta=5$,
  compared to the curve $m_t$. The inset is a zoom of the final times.}
\end{figure}

We have set the probabilities of the elementary processes to
$r=\delta t$, $p_r=p_l=\frac12(1-\delta t)$, and the lattice sizes to
$\delta t=0.01$ and $\delta x=0.1$; with this choice of parameters, the BRW
is a discretized version of the BBM. A realization of this
conditioned BRW
is displayed in Fig.~\ref{fig:realization}.

\begin{figure}[ht]
\includegraphics[width=\columnwidth]{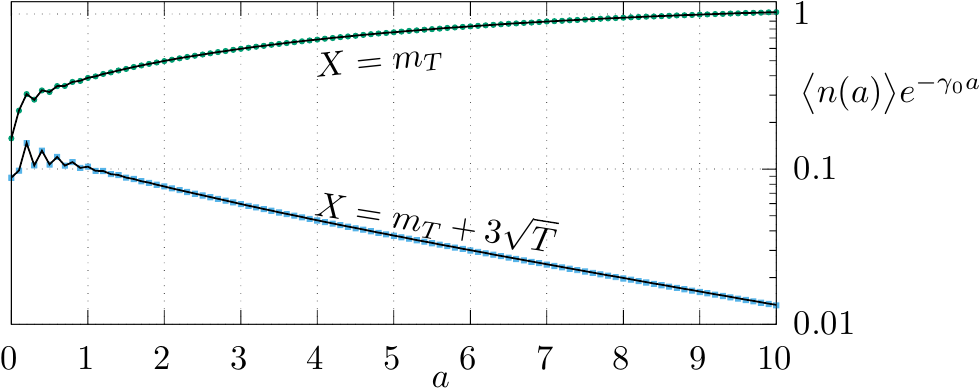}
\caption{\label{fig:mean}
 The expected number $\langle n(a)\rangle$ of particles at site $R_T-a$, multiplied by
 $e^{-\gamma_0 a}$ (where $\gamma_0=1.43195\cdots$ in the model used here),
 as a function of $a$, for $T=400$ and two values
of $X$. The dots are obtained from $3\,10^6$
realizations of the algorithm outlined in this letter.
The lines are from a numerical integration following the
methods of \cite{BrunetDerrida.2011}.
For $a=0$, we have removed 1 to the count of particles.}
\end{figure}

In order to validate our algorithm and its implementation,
we have measured
the expected number of
particles at  distance $a$ from the lead particle, because this quantity
can also be evaluated
from the formalism developed in~\cite{BrunetDerrida.2011} by
solving numerically an equation related to (\ref{FKPP2}). (See also the
discussion in the introduction).
The results displayed in Fig.~\ref{fig:mean} show a perfect
agreement between both methods
within statistical uncertainties.

\section{Continuous limit: conditioning the BBM}

The branching Brownian 
motion (BBM) is the continuous version of the BRW. The particles in a BBM
perform independent Brownian motions and branch with rate 1 (so that
during each infinitesimal time $\diffd t$, each particle 
is replaced by two particles with
probability $\diffd t$).

The method described in the previous section can be adapted to the BBM;
the goal is not necessarily to generate realizations, but to offer
a starting point to analytical studies of the tip
in typical and extreme events.

Introduce as before $u(t,x)=\P(R_t\ge x)$  as the probability that the rightmost
particle at time $t$ is on the right of $x$. It
satisfies the FKPP equation \eqref{FKPP} with initial condition
$u(0,x)=\indic{x\le0}$.
Introduce also red particles in the BBM, as in the BRW. The probability that
a particle is red is still $U(t,x)=u(T-t,X-x)$. We first consider the
probability that a particle, conditioned to be red, branches into two red
particles between $t$ and $t+\delta t$. The reasoning leading to
\eqref{redbrr} is still valid and gives, to leading order in $\delta t$,
\begin{equation}\label{redbrr2}
\P\Big(\tikzbranch {red}{red}\;\Big|\; \tikzpart\text{ is red}\Big)
=
\delta t\, U(t,x)+\mathcal O(\delta t^2).
\end{equation}
(Compare to \eqref{redbrr} with $r=\delta t$.)
Similarly, the conditional probability for branching into a red and a non-red is
\begin{equation*}
\P\Big(\tikzbranch {red}{non-red}\;\Big|\;\tikzpart\text{ is
red}\Big)=\delta t\, 2\big[1-U(t,x)\big]+\mathcal O(\delta t^2).
\end{equation*}
(Compare to \eqref{redbr}.)
We now turn to the motion of one single red particle during a time $\delta t\ll1$.
As branching occurs with small probability of order $\delta t$, we ignore this
possibility in the discussion below.
The probability that
a particle at $(t,x)$ moves during $\delta t$ by $\Delta
x\in[\epsilon,\epsilon+\diffd \epsilon]$ (which we write
for short $\Delta x\in\diffd \epsilon$) and is red is
\begin{equation*}
\P\Big(\Delta x\in\diffd \epsilon\et\tikzpart\text{ is red}\Big)
=\frac{e^{-\frac{\epsilon^2}{2\,\delta t}}}{\sqrt{2\pi \,\delta t}}\,\diffd
\epsilon \times U(t+\delta t,x+\epsilon).
\end{equation*}
Dividing by $U(t,x)$ gives $\P\big(\Delta x\in\diffd \epsilon\;|\;\tikzpart\text{ is
red}\big)$; then, multiplying by $\epsilon$ and integrating over $\epsilon$, we obtain
after changing variable $\epsilon=z\,\sqrt{\delta t}$ and expanding for small
$\delta t$,
\begin{align}
\big\langle \Delta x\;\big|\;\tikzpart\text{ is red}\big \rangle
&=\delta t\,\partial_x \ln U(t,x)+\mathcal O(\delta t^2).
\label{EDeltaxbis}
\end{align}
With \eqref{EDeltaxbis} and \eqref{redbrr2}, we thus obtain
the following result:

\smallskip

\emph{The trajectories of the particles in a BBM ending on the right of $X$
at time $T$, conditioned on the event that there is at least one of them,
is a BBM with a space- and time-dependent drift $\partial_x \ln U(t,x)$ and
a space- and time-dependent branching rate $U(t,x)$.}

\smallskip

If orange particles are needed, one checks that a red particle
branches out an orange particle at rate $2V_\Delta(t,x)$, that an orange
particle branches into two orange at rate $V_\Delta(t,x)$ and that orange
particles have a drift $\partial_x\ln V_\Delta(t,x)$.

\medskip

There is another way to construct the tree of red particles in the BBM.
Consider a particle at $(t,x)$ and call $(\tau_1,\xi_1)$ the time
and place of the next branching event (so that $\tau_1>t$).
One has, for $x_1\in\mathbb R$ and $t_1>t$,
$$\P(\tau_1\in\diffd t_1\et \xi_1\in\diffd x_1)
= e^{-(t_1-t)}\diffd
t_1\times\frac{e^{-\frac{(x_1-x)^2}{2(t_1-t)}}}{\sqrt{2\pi(t_1-t)}}\diffd x_1
.$$
For $t_1<T$, the probability that the particle that just branched
is furthermore red is
obtained by multiplying the right hand side by $U(t_1,x_1)[2-U(t_1,x_1)]$,
the probability that at least one of the two children is red.
Then, dividing by $U(t,x)$  we obtain
the conditional probability
\begin{multline}\label{x1t1}
\P(\tau_1\in\diffd t_1\et \xi_1\in\diffd x_1\;|\; \tikzpart\text{ is red})
 = e^{-(t_1-t)}\diffd t_1\\
\times\frac{e^{-\frac{(x_1-x)^2}{2(t_1-t)}}}{\sqrt{2\pi(t_1-t)}}\diffd x_1
\times\frac{U(t_1,x_1)[2-U(t_1,x_1)]}{U(t,x)}.
\end{multline}
Note that this probability is not normalized: the integral of \eqref{x1t1}
on $x_1\in\mathbb R$ and on $t_1\in[t,T]$ is smaller than 1,
and the remaining probability corresponds to the event that the next branching
occurs after the time horizon $T$. In that case, the trajectory up to
time $T$ of the red particle is simply a Brownian motion (no branching)
conditioned to finish on the right of $X$.

With \eqref{x1t1} one can draw the coordinates $(\tau_1,\xi_1)$ of the next branching
event. The
trajectory between $t$ and $\tau_1$ is then a Brownian motion conditioned to
be at position $(\tau_1,\xi_1)$. It remains to determine the type of branching at
time $\tau_1$. With no conditioning, the probability to branch into two red is
$U^2$, writing for short  $U$ instead of $U(\tau_1,\xi_1)$, and the probability to branch into one red and one non-red
is $2U(1-U)$. Then, given that the branching  particle is red,
the probability that it branches
into two red is $U/(2-U)$. 
With the complementary probability, only one red particle
remains. In either case, the algorithm is restarted from $(\tau_1,\xi_1)$.

\section{Variant: Fixing the exact position of the rightmost particle}

We briefly present a variant of our algorithm. 
The idea
is to condition the red particles at time $T$ to be \emph{exactly} at
position $X$, rather than in $[X,\infty)$. In the BRW, the probability for
a particle at $(t,x)$ to reach $(T,X)$ is
$$\tilde U(t,x):=u(T-t,X-x)-u(T-t,X-x+\delta x).$$
Compare to \eqref{defU}.
Then, following the same argument as above, the
evolution probabilities for these ``new red''
particles are given by~\eqref{redjr},
\eqref{redjl}, \eqref{redbrr} and \eqref{redbro}
with $U$ replaced by $\tilde U$ and by \eqref{redbrb} with the
two $U$ outside the square brackets replaced by $\tilde U$.
With these new equations, one can follow the ``new red'' (and the orange)
particles.

This variant is a bit more difficult to implement correctly, because it
is harder to obtain a good numerical precision for $\tilde U$ than for $U$.
Its advantage is that it allows to generate unbiased realizations of all
the
particles on the right of $R_T-\Delta$ in a BRW. To do this, for each
realization, we first draw the value of $R_T$; as we need anyway to
compute $u(T,x)=\P(R_T\ge x)$, this operation is easy. Then, we run the
variant of our algorithm with $X=R_T$ to generate the ``new red''
particles (ending at $X$) and the orange particles (ending in $[X-\Delta,X)$~).
The resulting particles form an unbiased realization of the tip of
the BRW.

For the BBM, the same variant can also be used; the probability
of ending in $\diffd X$ is $\partial_x U(t,x)\,\diffd X$, and one finds
that the ``new red'' particle follows a Brownian motion with  drift
$\partial_x\ln[\partial_x U(t,x)]$, compare to \eqref{EDeltaxbis}.
This particle branches out particles
conditioned to end to the left of $X$ 
with rate $2\big[1-U(t,x)\big]$. (More precisely, it branches orange
particles with rate $2 V_{\Delta}(t,x)$ and blue particles with rate
$2\big[1-U(t,x)- V_{\Delta}(t,x)\big]=2\big[1-U(t,x+\Delta)\big]$). It cannot branch
into two ``new red'' particles, because there is a probability zero that
a second particle ends up exactly at position $X$. This description of the
BBM, with exactly one marked particle branching BBMs conditioned to not
overtake it, is the same as the spine description~\cite{HardyHarris.2009}.

\section{Conclusion and outlook}

We have presented a simple algorithm to generate
only the tip (the rightmost particles) in realizations of BRW in which the
rightmost particle is
constrained to be on the right of an arbitrary position $X$ at an
arbitrary time $T$ or, in a variant, to be exactly at $X$. We have validated it by
comparing Monte Carlo calculations obtained with this
algorithm to predictions obtained by a different method.

When $X$ is large compared to the expected position $m_T$ of the rightmost
particle at time $T$, our algorithm allows to study rare realizations.
When $X$ is close to $m_T$, it allows to generate more typical realizations.

\begin{figure}[ht]
\includegraphics[width=\columnwidth]{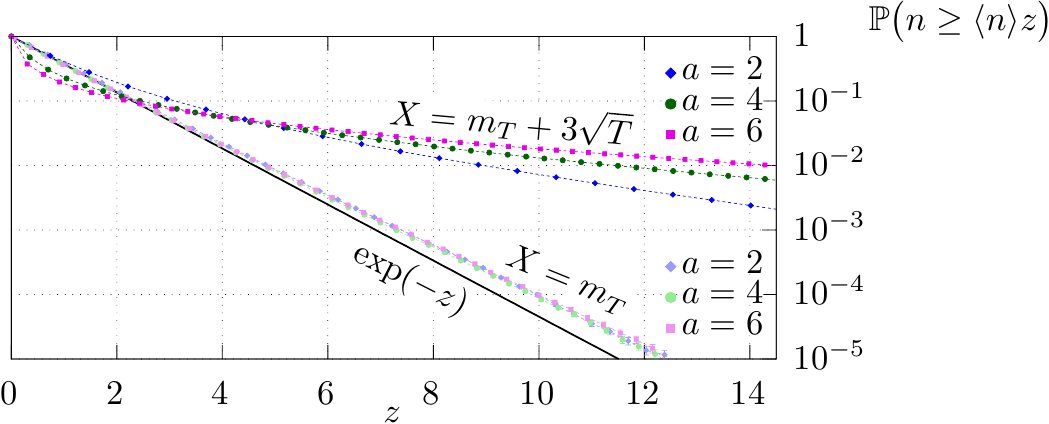}
\caption{\label{fig:dist}
  Rescaled tail distribution $\P\big(n\ge \langle n\rangle z\big)$
  as a function of $z$ for the number $n$ of particles at position
$R_T-a$ with $T=400$, for two values of $X$ and three values of $a$.
  These probabilities were measured from the same realizations as for 
  Fig.~\ref{fig:mean}.
The distribution of $n$ is roughly exponential for $X=m_T$, but
 it exhibits a much fatter tail when $X=m_T+3\sqrt T$.}
\end{figure}

Our algorithm enables the study of observables of the tip region of the BRW for
which no other method is available to date.
For example, we have measured numerically the distribution
of the number of particles at distance $a$ to the left of
the rightmost, in typical and rare realizations,
see Fig.~\ref{fig:dist}, and this will allow to check a recent
heuristic calculation in~\cite{MuellerMunier.2019}.

 Among the further developments made possible
  by this algorithm, we intend to investigate
  observables such as the distribution of the genealogical tree of the
particles in the tip \cite{ArguinBovierKistler.2011,
  DerridaMottishaw.2016}.

While our main focus has been the BRW, we have also given a theoretical
description of the conditioned BBM which may be useful to give
a mathematical description of the tip along lines similar to
\cite{AidekonBerestyckiBrunetShi.2012}.

\medskip

\begin{acknowledgments}
  We thank the CPHT computer support team for the maintenance of the
  cluster ``hopper'' of the PHYMATH mesocenter (\'Ecole polytechnique and CNRS)
  on which the numerical calculations presented here were performed.
\end{acknowledgments}

\end{document}